\documentclass[prx,amsmath,groupedaddress,floatfix,twocolumn]{revtex4-1}

\usepackage{graphicx}
\usepackage{amssymb}
\usepackage{longtable}
\usepackage{rotating}
\usepackage{array}
\usepackage{multirow}
\usepackage{makecell}
\usepackage{booktabs}
\usepackage{xcolor}

\begin{document}


\title{Meta-screening and permanence of polar distortion in
  metallized ferroelectrics}

\author{Hong Jian Zhao,$^{1}$ Alessio Filippetti,$^{2,3}$ Carlos
  Escorihuela-Sayalero,$^{1}$ Pietro Delugas,$^{4}$ Enric
  Canadell,$^{5}$ L. Bellaiche,$^{6}$ Vincenzo Fiorentini$^{2,3}$ and
  Jorge \'I\~niguez$^{1}$}

\affiliation{
  $^{1}$ \mbox{Materials Research and Technology
  Department, Luxembourg Institute of Science and Technology (LIST),} 
  \mbox{Avenue des Hauts-Fourneaux 5, L-4362 Esch/Alzette,
  Luxembourg}\\
  $^{2}$ \mbox{Dipartimento di Fisica, Universit\`a di Cagliari, Cittadella
  Universitaria, I-09042 Monserrato (CA), Italy}\\
  $^{3}$ \mbox{CNR-IOM Cagliari, 
  Cittadella Universitaria, I-09042 Monserrato (CA), Italy}
  $^{4}$\mbox{Scuola Internazionale Superiore di Studi Avanzati, Via Bonomea
  265, I-34136 Trieste, Italy}\\
  $^{5}$ \mbox{Institut de Ci\`encia de Materials de Barcelona (ICMAB-CSIC),
  Campus UAB, 08193 Bellaterra, Spain}\\
  $^{6}$ \mbox{Physics Department and Institute for Nanoscience and
  Engineering, University of Arkansas,} Fayetteville, Arkansas 72701,
  USA\\
}%

\begin{abstract}
Ferroelectric materials are characterized by a spontaneous polar
distortion. The behavior of such distortions in the presence of free
charge is the key to the physics of metallized ferroelectrics in
particular, and of structurally-polar metals more generally. Using
first-principles simulations, here we show that a polar distortion
resists metallization and the attendant suppression of long-range
dipolar interactions in the vast majority of a sample of 11
representative ferroelectrics. We identify a {\em
  meta-screening} effect, occurring in the doped compounds as a
consequence of the charge rearrangements associated to electrostatic
screening, as the main factor determining the survival of a
non-centrosymmetric phase. Our findings advance greatly our
understanding of the essentials of structurally-polar metals, and
offer guidelines on the behavior of ferroelectrics upon field-effect
charge injection or proximity to conductive device elements.
\end{abstract}


\maketitle

\section{\label{sec:intro}Introduction}

In many materials, spontaneous structural distortions occur that break
the inversion symmetry of a parent centrosymmetric (CS)
structure. These are usually named polar distortions (PDs), since they
enable the existence of non-zero polar-vector observables, such as
spontaneous electric polarization. Ferroelectrics (FEs) display just
such a PD and consequently possess a spontaneous polarization. By
definition \cite{lines-book1977}, in a FE polarization must be
switchable by an external field (non-switchable polarized materials do
exist, named pyroelectrics \cite{ambacher02}). Because of this
requirement, ferroelectrics should be insulators or semiconductors, as
opposed to metals, so that they can be acted upon with an external
bias. However, it is not {\em a priori} obvious that the insulating character
itself is necessary for a PD to occur: could it not \cite{anderson65}
happen in a metal?

Our general understanding of basic ferroelectric phenomena -- largely
based on empirical \cite{lines-book1977,strukov-book1998} and early
first-principles \cite{cohen92,posternak94,zhong94a,rabe-book2007}
studies of perovskite oxides such as BaTiO$_{3}$, PbTiO$_{3}$, or
KNbO$_{3}$, centers on the role of electrostatic dipole-dipole
couplings as the driving force of the long-range polar order. As a
result, free carriers and the attendant electrostatic screening are
usually regarded as incompatible with the existence of PDs. Hence, at
least among perovskite oxides \cite{benedek16}, non-centrosymmetric
metals (NCSMs) are usually deemed exotic. This viewpoint has been
supported by theoretical work on BaTiO$_{3}$ \cite{iwazaki12,wang12},
whose results seem to be taken as a general rule.

NCSMs are currently a hot topic for obvious reasons of fundamental
understanding, but also because of the possible occurrence of quantum
phenomena in the context of superconductivity
\cite{yildirim13,bauer-book2012}, and of course their technological
relevance to devices involving conductive and FE elements. Indeed,
considerable efforts \cite{shi13,filippetti16,puggioni14,kim16} are
currently focused on the experimental discovery and first-principles
prediction of NCSM compounds, and are yielding experimental
\cite{shi13}, and very recently theoretical
\cite{filippetti16,benedek16,he16a,he16b,shimada17}, results that
question the common wisdom that metallization is incompatible with the
occurrence of a PD. For example, first-principles studies have
recently suggested that the PD of materials like PbTiO$_{3}$ and
BiFeO$_{3}$ is not strongly affected by the presence of free carriers
\cite{he16a,shimada17,he16b}. Further, some of us took advantage of
the chemical origin of ferroelectricity in Bi-based compounds to
predict a {\em switchable} polar order in Bi$_{5}$Ti$_{5}$O$_{17}$, a
layered perovskite that is metallic \cite{filippetti16}.  A careful
examination and rationalization of the compatibility between PDs and
free carriers is thus certainly warranted, both to buttress our
fundamental understanding and to suggest practical routes to obtain
NCSMs, for example by the metallization of a known ferroelectric
compound (e.g., by suitable chemical doping or field-effect charge
injection).

Here we analyze the effect of doping on PDs by studying from first
principles a collection of diverse and representative FE materials. We
find that the PD coexists with metallicity in most of the considered
compounds. We discuss the atomistic interactions responsible for the
observed behaviors, revealing a largely universal {\em meta-screening}
effect that favors polar distortions upon doping. As a by-product of
our work, we come up with obvious prescriptions to obtain FE materials that
should yield non-centrosymmetric metals upon doping. Other
implications of our results, e.g., as regards hyperferroelectric
effects, are also briefly discussed.

\section{\label{sec:results}Results and discussion}

We consider a total of 11 ferroelectric compounds that represent
different families owing their FE order to different physical and
chemical mechanisms. More specifically, we have LiNbO$_{3}$ (LNO),
several perovskites (BaTiO$_{3}$ or BTO, KNbO$_{3}$ or KNO,
PbTiO$_{3}$ or PTO, BiFeO$_{3}$ or BFO, BaMnO$_{3}$ or BMO, and
BiAlO$_{3}$ or BAO), and layered perovskites (La$_{2}$Ti$_{2}$O$_{7}$
or LTO227, Sr$_{2}$Nb$_{2}$O$_{7}$ or SNO227, and
Ca$_{3}$Ti$_{2}$O$_{7}$ or CTO327), and a (001)-oriented superlattice
formed by LaFeO$_{3}$ and YFeO$_{3}$ perovskite layers that are one
unit cell thick (LFO/YFO). Beyond these, we also consider other
paraelectric perovskite compounds (LaAlO$_{3}$ or LAO), and even
metals (Cr and V) and Zintl semiconductors (KSnSb or KSS), to run
additional calculations that aid our discussion. Most of our
calculations take the ground state structure of these materials, which
in all cases is known from the literature, as a starting point to
study their behavior upon doping. In a few cases we consider (or
identify) additional phases that are stabilized upon doping, and which
we introduce in due course. Further details on our calculations are in
Appendix~\ref{sec:methods}.

\subsection{Polar distortions under doping} 

We begin by discussing the behavior of PDs in our sample of FE
compounds as a function of doping.  We adopt the convention that a
positive carrier density $\rho_{\rm free}$ corresponds to extra
electrons (i.e., $n$ doping), while negative $\rho_{\rm free}$ values
indicate hole ($p$) doping. We relax all structures as a function of
carrier concentration, and monitor the evolution of the PD normalized
to its value in the undoped case (see Appendix~\ref{sec:methods} for
details).

\begin{figure}
\includegraphics[width=0.95\columnwidth]{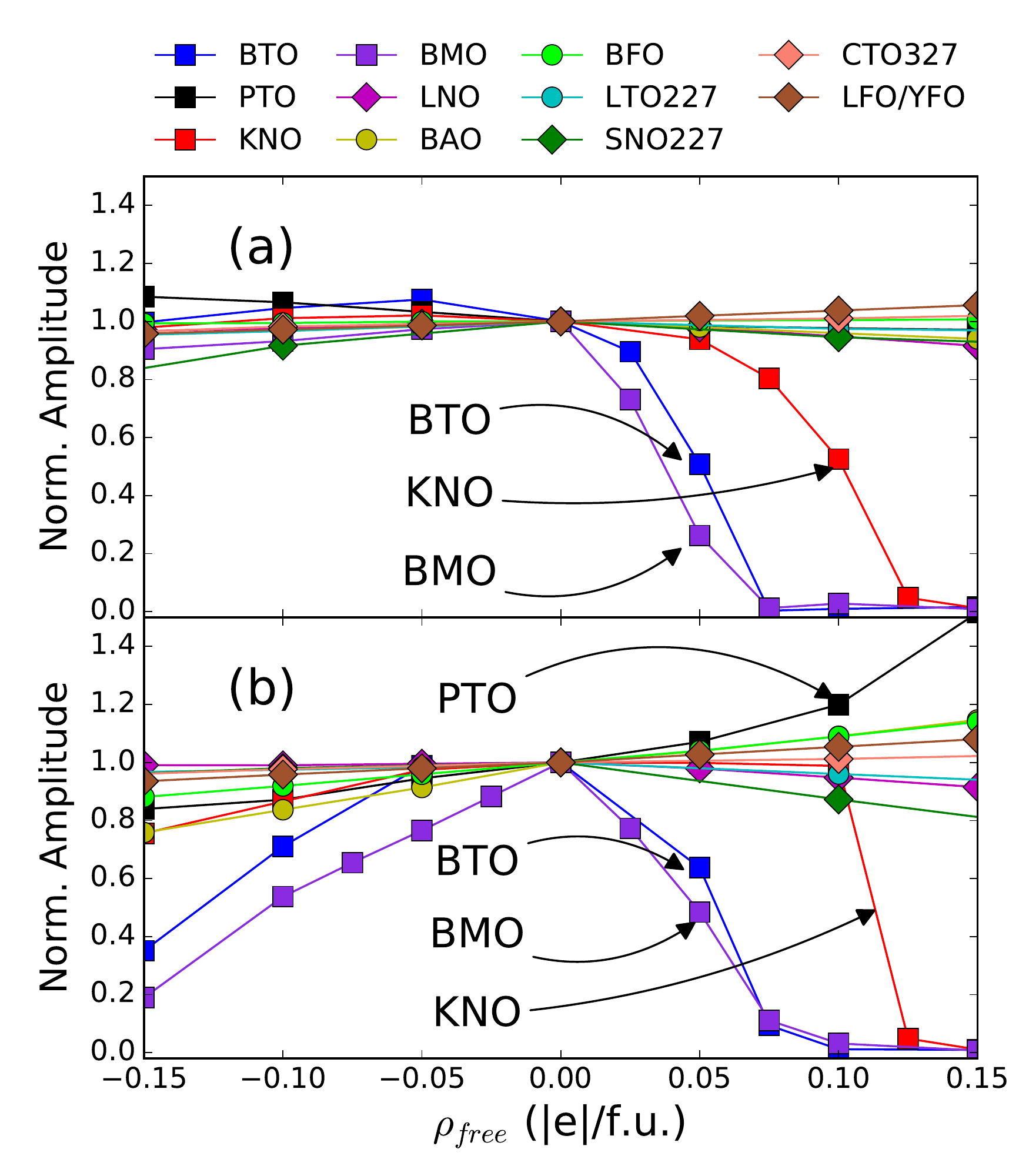}
  \caption{Calculated magnitude of the polar distortion as a function
    of doping with electrons ($\rho_{\rm free} > 0$) and holes
    ($\rho_{\rm free} < 0$). (a) Shows the results when we
    impose the volume of the undoped solution be preserved upon doing,
    while (b) shows the results when the volume is allowed to
    relax. The cell shape is always allowed to relax. The polar
    distortion is quantified as described in the text, and normalized,
    for each considered compound, to its value in the undoped
    case. Note that, for perovskite oxides with a five-atom formula unit
    (henceforth f.u.), $\rho_{\rm free} = 0.1$~$|e|$/f.u. corresponds
    to a charge density of about $1.5\times 10^{21}$~cm$^{-3}$. $e$ is
    the electron charge.}
  \label{fig:fe-vs-doping}
\end{figure}

In Fig.~\ref{fig:fe-vs-doping}(a), we present the results obtained
under the constraint that the unit-cell volume be fixed and equal to
the value obtained in the undoped case. In
Fig.~\ref{fig:fe-vs-doping}(b), we show instead the corresponding data
when a full volume relaxation is permitted.
Figures~\ref{fig:fe-vs-doping}(a) and \ref{fig:fe-vs-doping}(b) display
the same qualitative behavior; the distinction is relevant for reasons
to be discussed below.

Figure~\ref{fig:fe-vs-doping} yields one clear main message: the PDs
survives metallization in the vast majority of the considered FE
compounds. The PD is unaffected or reinforced in materials in which
ferroelectricity is mainly driven by chemical or steric effects (as in
PbTiO$_{3}$, BiFeO$_{3}$, BiAlO$_{3}$, and LiNbO$_{3}$), caused by a
particular lattice topology or geometry (as in La$_{2}$Ti$_{2}$O$_{7}$
and Sr$_{2}$Ti$_{2}$O$_{7}$ \cite{lopez-perez11}), or an improper
effect triggered by a different primary order parameter (as in
Ca$_{3}$Ti$_{2}$O$_{7}$ \cite{oh15,benedek11} and
LaFeO$_{3}$/YFeO$_{3}$ \cite{zanolli13,rondinelli12} superlattices).
In fact, in our doping range, the PD disappears only for BTO, BMO, and
KNO under $n$ doping, and even then, it does take quite some free
charge (well above 10$^{21}$~cm$^{-3}$) to kill it.

\begin{figure}
\includegraphics[width=0.95\columnwidth]{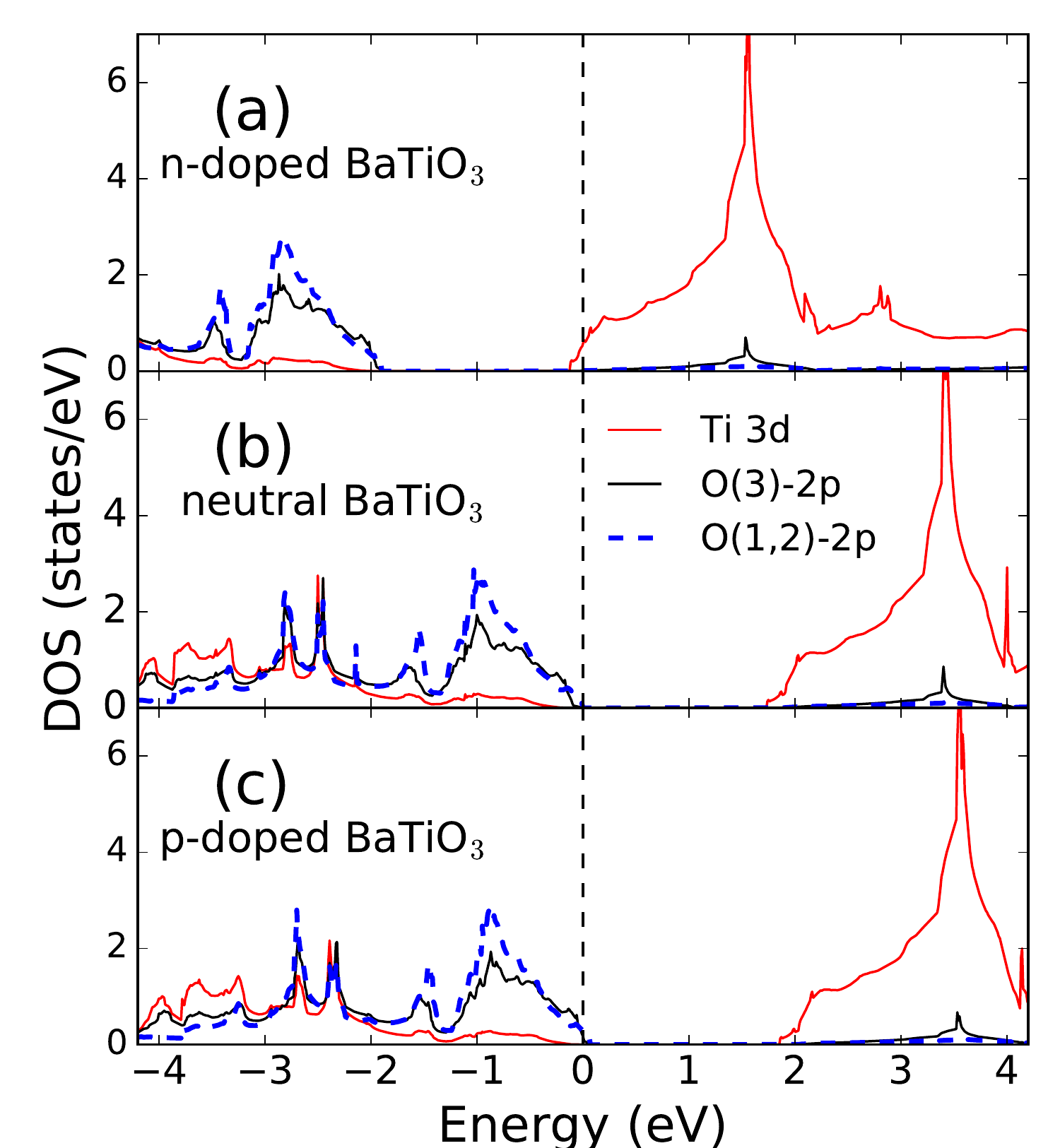}
  \caption{Partial density of states of BaTiO$_{3}$ under doping. We
    show the results for $n$ doping [$\rho_{\rm free} =
      0.05$~$|e|$/f.u., panel~(a)], the undoped case [panel~(b)], and
    $p$ doping [$\rho_{\rm free} = -0.05$~$|e|$/f.u., panel~(c)]. The
    Fermi level is chosen as zero of energy in all cases.}
  \label{fig:dos}
\end{figure}

In our description (see also Appendix~\ref{sec:methods}) of doping,
charge localization, e.g., into narrow gap states, is excluded since we
work with perfect crystals, the periodic unit being that of the
undoped compound. Hence, the doping charges occupy itinerant Bloch
states at the conduction band bottom (electrons) or valence band top
(holes), as illustrated by the density of states of BaTiO$_{3}$ in
Fig.~\ref{fig:dos}, which is representative of all materials.

\subsection{Screening and interactions under doping}

To better understand how doping affects the PD, we inspect the effect
of the carriers on the relevant interatomic interactions. We
specifically analyze the behavior of BTO, BMO, PTO, and BFO, four
perovskites that share some similarities, but also present key
differences. For example, in both BTO and BMO the PD is mainly driven
by the off centering of the {\sl B} cations, and is known to rely
strongly on dipole-dipole interactions
\cite{zhong94a,zhong95a,bhattacharjee09}. However, Ti$^{4+}$ has a
3$d^{0}$ electronic configuration, while Mn$^{4+}$ presents a 3$d^{3}$
state; hence, the doping electrons and holes occupy different types of
orbitals in these two compounds. On the other hand, BFO is a material
in which the (very large) PD is driven by the {\sl A} cation and has a
widely accepted chemical origin (Bi$^{3+}$'s lone pair)
\cite{ravindran06,dieguez11}. Finally, PTO is a material that shares
features of BTO (Ti$^{4+}$ in a 3$d^{0}$ state, with large
dipole-dipole interactions) and BFO (Pb$^{2+}$'s lone pair).

\begin{figure}
\includegraphics[width=0.95\columnwidth]{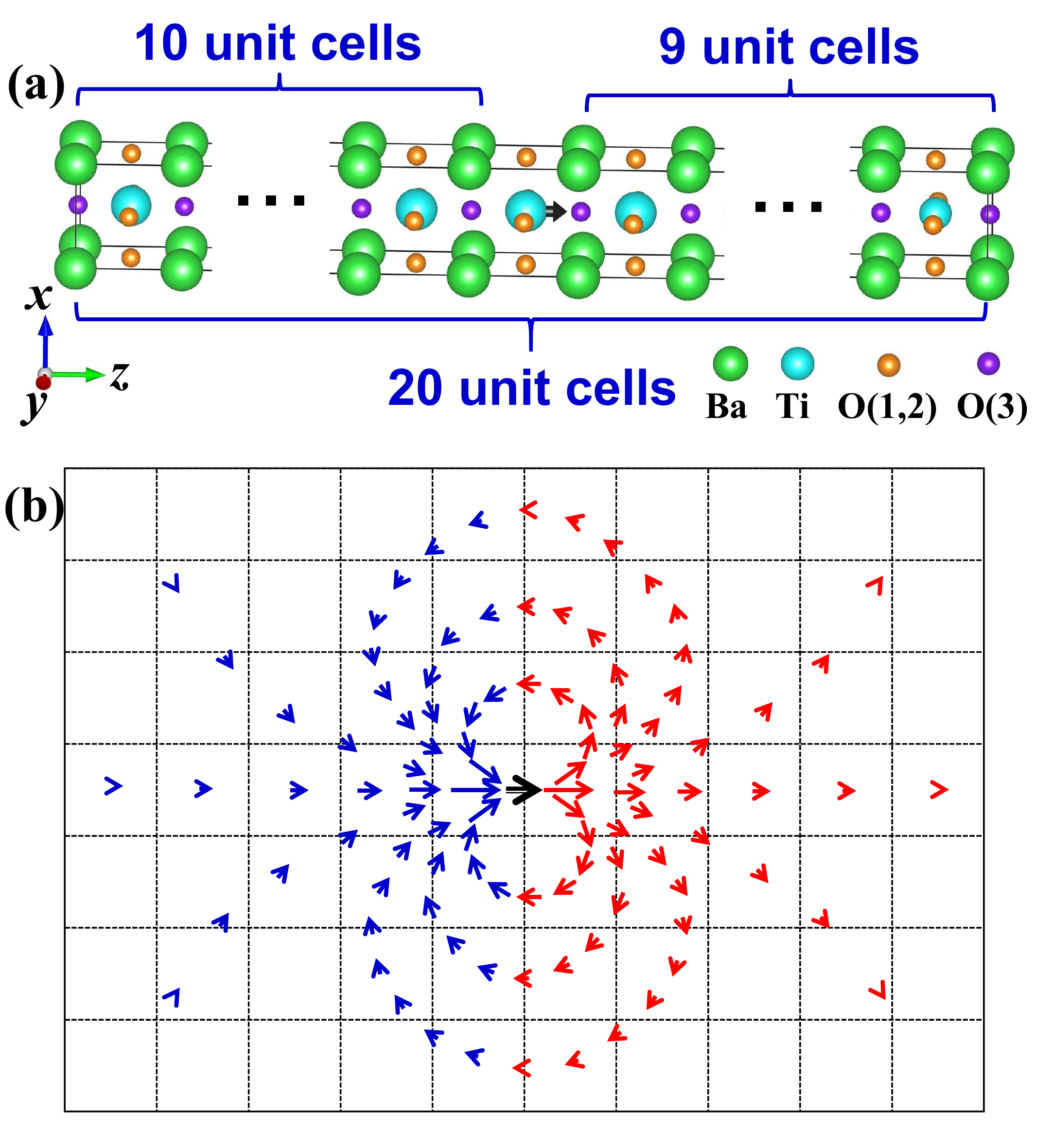}
  \caption{(a) Shows a sketch of the supercell used to
    investigate the response of doped BaTiO$_{3}$ to a plane of
    dipoles created by displacing Ti atoms along $z$. Atoms types,
    coordinates, and other elements mentioned in the text are
    indicated. In (b) we sketch the dipole field created by a
    displaced Ti atom, to stress the simultaneous occurrence of
    parallel longitudinal interactions and anti-parallel lateral
    ones.}
  \label{fig:supercell}
\end{figure}

\subsubsection{BaTiO$_{3}$: raw results}

\begin{figure}
\includegraphics[width=0.95\columnwidth]{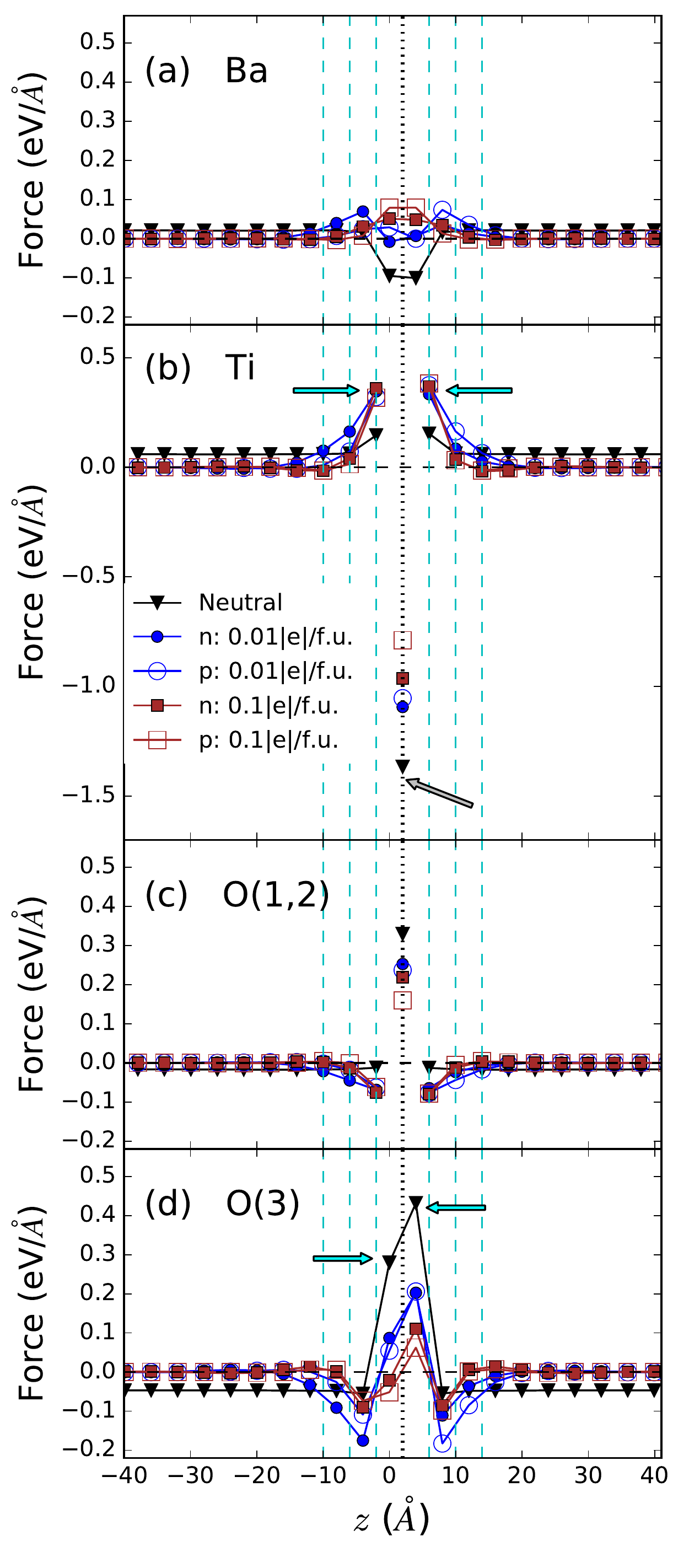}
  \caption{Forces occurring in response to the plane of dipoles in
    BaTiO$_{3}$. We create the dipole plane by displacing along $z$
    the Ti atoms located at $z \approx 2$~\AA, marked with a black
    dotted line. Results are shown for different doping levels, and we
    mark with dashed lines the TiO$_{2}$ planes within the regions in
    which screening charges accumulate (see text). We show the forces
    acting on Ba [panel~(a)], Ti [panel~(b)], O(1) and O(2)
    [panel~(c)], and O(3) [panel~(d)] atoms. For all atoms, the $x$
    and $y$ components of the force are zero by symmetry; hence, we
    only show the $z$ component. We use arrows to highlight forces
    associated to especially important interactions (see text). Note
    that we use lines to guide the eye, except for the data points at
    $z \approx 2$~\AA\ in (b) and (c), to aid visibility.}
  \label{fig:forces-BTO}
\end{figure}

\begin{figure}
\includegraphics[width=0.95\columnwidth]{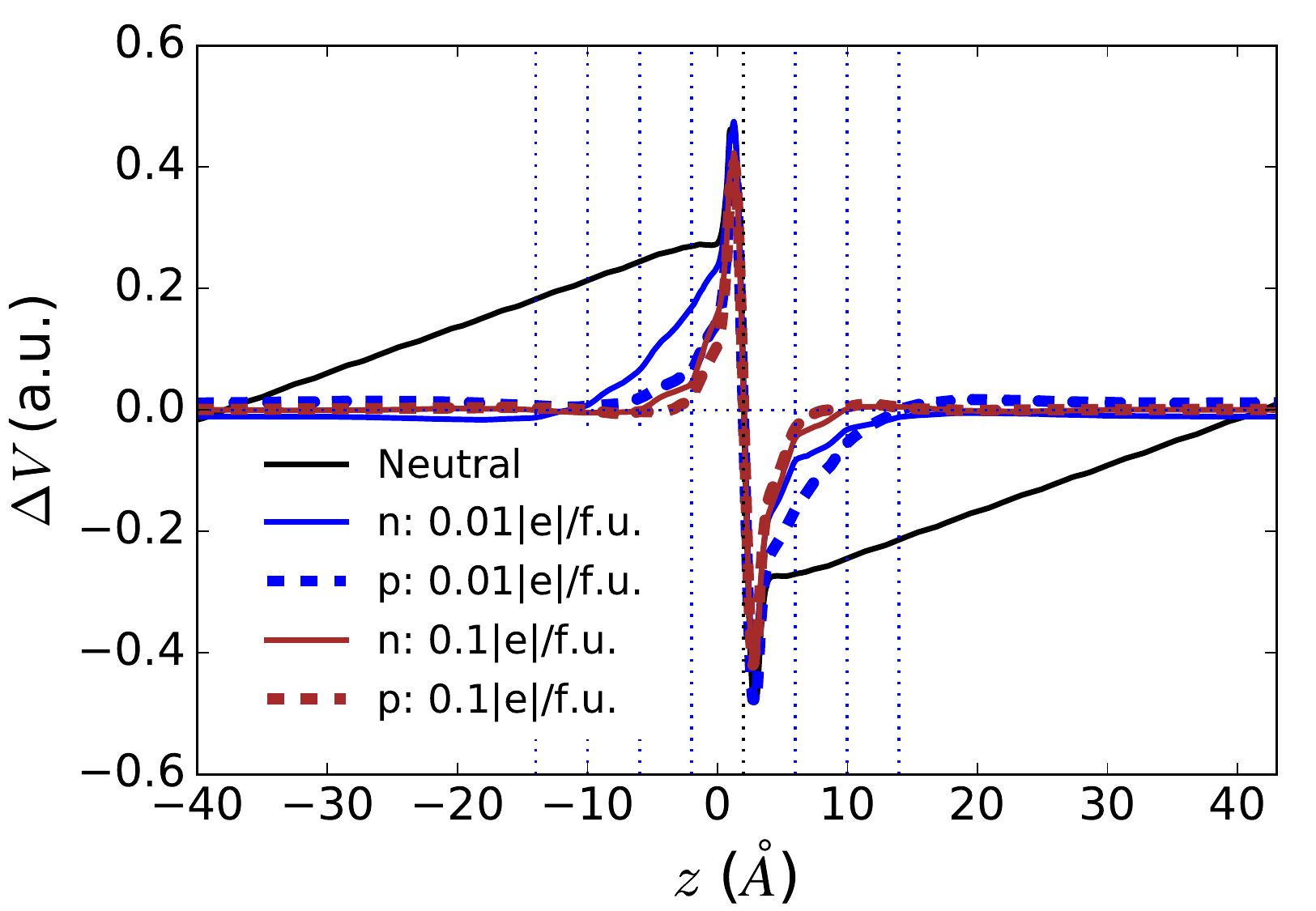}
  \caption{Changes in the electrostatic potential, as computed for
    BaTiO$_{3}$ under different doping levels, and associated to the
    Ti displacement that creates a plane of dipoles. The cases shown
    correspond to those of Fig.~\ref{fig:forces-BTO}. We plot the
    difference potential $\Delta V(z) = V_{\rm dist}(z) - V_{\rm
      cubic}(z)$, obtained by comparing the result for the ideal cubic
    lattice ($V_{\rm cubic}$) with the one obtained in presence of the
    Ti distortion ($V_{\rm dist}$). Relevant TiO$_{2}$ planes are
    marked as in Fig.~\ref{fig:forces-BTO}. To plot these potential
    differences, we perform an in-plane average of the results from
    our simulations, but no average along the $z$ direction.}
  \label{fig:potential-BTO}
\end{figure}

\begin{figure}
\includegraphics[width=0.95\columnwidth]{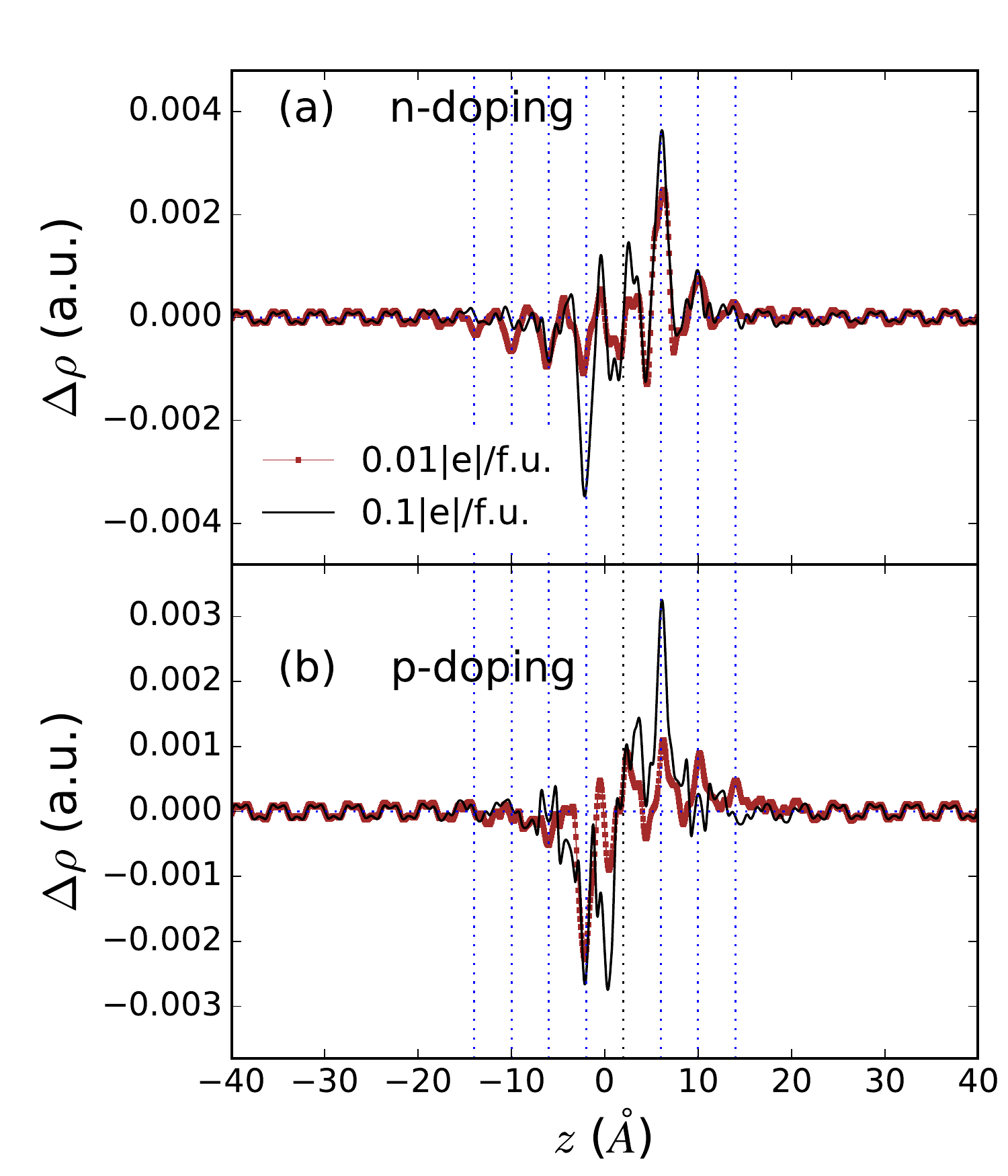}
  \caption{Electronic rearrangement associated to the electrostatic
    screening in BaTiO$_{3}$, as occurring in our supercell
    simulations imposing a plane of dipoles, for different doping
    levels. The cases shown correspond to those of
    Fig.~\ref{fig:forces-BTO}. We plot the difference density $\Delta
    \rho (z) = \rho_{\rm dist}(z) - \rho_{\rm cubic}(z)$, obtained by
    comparing the result for the ideal cubic lattice ($\rho_{\rm
      cubic}$) with the one obtained in presence of the Ti distortion
    that creates the plane of dipoles ($\rho_{\rm dist}$). Relevant
    TiO$_{2}$ planes are marked as in Fig.~\ref{fig:forces-BTO}. To
    plot these electronic density differences, we perform a
    macroscopic average (using a window of 1.9~\AA\ along the $z$
    direction) of the raw results from our simulations.}
  \label{fig:charge-BTO}
\end{figure}

We first focus on BTO, the material where the PD is the least robust
of all. To visualize the interactions responsible for the FE
instability of BTO, we run the following simulations. We consider the
long supercell sketched in Fig.~\ref{fig:supercell}(a), which
comprises 1$\times$1$\times$20 elemental 5-atom units, with the atoms
in their high-symmetry (cubic phase) positions. Then, we displace by
0.05~\AA\ along $z$ the Ti atom in the first cell, noting that,
because we work with a periodically-repeated supercell, this amounts
to creating an array of $xy$ planes of $z$-polarized dipoles,
separated by 19 unit cells (about 76~\AA) from each other. Then we
compute the forces, considering the undoped case as well as
representative doping values. The results are summarized in
Figs.~\ref{fig:forces-BTO}, \ref{fig:potential-BTO}, and
\ref{fig:charge-BTO}.

\subsubsection{Undoped BaTiO$_3$}

In the undoped case, we find that the force acting on the displaced Ti
atom is large and negative. This is a restoring force resulting from
two types of interactions: one, short-range repulsive coupling
between the Ti and its neighboring oxygens; two, long-range
interactions between dipoles within the $z \approx 2$~\AA\ plane, as
well as with their periodic images. As sketched in
Fig.~\ref{fig:supercell}, the lateral interactions between the dipoles
in a given $xy$ plane favor an antipolar order, i.e., they add to the
restoring force acting on our displaced Ti. In particular, by
performing the corresponding Ewald sum, we estimate this dipole-dipole
contribution to be about $-$0.35~eV/\AA\ in the present case, which is
about 25~\% of the total force of $-$1.37~eV/\AA\ obtained in our
calculation. (The dominant interactions are those between dipoles in
the same plane; the coupling with periodic-image dipole planes is very
small.)

If we now move to the two apical oxygens [labeled O(3) in
  Fig.~\ref{fig:supercell}] that lie closest to the displaced Ti, we
find relatively large and positive forces acting on them. If we try to
understand such forces as the result of short- and long-range
interactions, it becomes apparent that they must be dominated by the
former kind. Note that the positive dipoles created by the plane of
displaced Ti atoms yield a net positive electric field on these O(3)
oxygens, which should result in {\em negative} dipole-dipole
forces. [The relevant dynamical charges are 7.73~$|e|$ for Ti and
$-$6.15~$|e|$ for O(3).] Hence, the computed positive forces must thus
be the result of a stronger and repulsive short-range interaction
between the Ti and O(3) atoms; this interaction can be seen as tending
to preserve an optimal Ti--O(3) distance. Note also that the force
computed for the O(3) on the left of the displaced Ti is different
from that of the O(3) on the right; this is quite natural, as these
two O(3) atoms are not related by symmetry in the distorted
configuration; in fact, this difference reflects anharmonic
interactions that have an effect even though the considered
displacement of the Ti atom (0.05~\AA) is relatively small.

As regards the equatorial oxygens [O(1) and O(2)] that are nearest
neighbors from the displaced Ti, the obtained positive forces are not
a surprise, as both short-range [which will tend to preserve the
optimum Ti--O(1) distance in the cubic phase] and long-range [the
dipole field in the $xy$ dipole plane is negative] interactions give a
positive contribution. [In this case, the relevant dynamical charge
  for O(1) and O(2) is about $-$2.15~$|e|$.] As regards the Ba atoms,
we obtain relative small forces that we do not discuss here.

Interestingly, none of the forces just mentioned, which act on atoms
close to the dipole plane, tend to stabilize the polar
distortion. Indeed, they are all restoring forces, and it seems safe
to interpret them as dominated by short-range (repulsive) couplings
favoring the high-symmetry cubic structure. (Short-range interactions
are indeed often mentioned in the literature as detrimental to
ferroelectricity in BTO \cite{cohen92}.) However, the situation
changes drastically for atoms far from the dipole plane. For those, we
obtain finite forces saturating to a non-zero value at around
8~\AA\ from the displaced Ti: in that region, we observe positive
forces of about 0.06 and 0.02~eV/\AA\ acting on the Ti and Ba
atoms, respectively; and negative forces of about $-$0.02~eV/\AA\ and
$-$0.05~eV/\AA, respectively, acting on the O(1,2) and O(3)
anions. Such forces are the result of the quasi-homogeneous field that
the $xy$ dipole planes create in the intermediate region of the
supercell; as shown in Supplemental Material (Note 1 and Fig. 1) \cite{SupMat}, they can be
easily recovered from the potential (Fig.~\ref{fig:potential-BTO}) and
dynamical charges obtained from our simulations. (By performing the
corresponding Ewald sums \cite{allen-book1989} for our periodic planes
of spaced dipoles, we checked explicitly that, for the situation here
considered, a nearly constant field must indeed appear in the
intermediate regions. As the separation between dipole planes
increases, the field develops small spatial inhomogeneities and
eventually decays to zero away from the dipole planes.) These
dipole-dipole forces push the cations and anions to move against each
other, and thus tend to stabilize a PD. Hence, this is a manifestation
of the dipole-dipole interactions responsible for the PD of
ferroelectrics like BTO. From a related perspective, since there is no
free charge, the equilibrium state of the material should satisfy the
Maxwell relation for the electric displacement field $\nabla\cdot{\bf
  D}=\rho_{\rm free}=0$. Thus the computed forces in the intermediate
regions capture the response of the compound aiming at an homogeneous
state of constant $D_{z}$ when a dipole plane is created.

\subsubsection{Doped BaTiO$_3$: electrostatic screening}

Let us now discuss the results obtained under doping. One obvious
difference with the undoped case is that the forces vanish in the
regions away from the dipole plane. Correspondingly, as shown in
Fig.~\ref{fig:potential-BTO}, the computed potential is flat in those
areas. Hence, as expected, the presence of dopants, positive or
negative, renders a metallic system and permits the screening of the
dipole-dipole interactions. Naturally, this effect goes against the
onset of a PD.

We can appreciate how the screening comes about by comparing the DFT
results for the non-polar (cubic) and polar (Ti-displaced) structures,
as shown in Fig.~\ref{fig:charge-BTO}. For example, our results for
$n$ doping show that an excess of electrons appears in a region within
8~\AA\ to the right of the $xy$ dipole plane, while an excess of holes
occurs in a region of about 12~\AA\ on the left side.

The fact that these two regions are not symmetric makes physical
sense: In the cubic structure, the $n$-dopants occupy the Ti-$3d$
levels, and distribute homogeneously throughout the supercell. Upon
displacement of the Ti atom at $z \approx 2$~\AA, we essentially have
a transfer of mobile electrons from the Ti's on the left of the dipole
plane to the Ti's on the right side of it. Since the doping level is
low, the amount of mobile electrons available in the left-side Ti's is
small, and a relatively large number of atoms are required to provide
sufficient charges; in contrast, there are plenty of empty 3$d$
orbitals in the Ti's on the right, and the excess electrons can be
accommodated in a relatively small number of atoms. In the case of
$p$-doping [Fig.~\ref{fig:charge-BTO}(b)] we observe the same kind of
electron depletion (on the left) and accumulation (on the right), and
a similarly efficient electrostatic screening
(Fig.~\ref{fig:forces-BTO}); yet, the details are different,
reflecting the different orbitals involved in the charge
redistribution. Indeed, in this case the left-side electron donors are
O-2$p$ orbitals, and it is also O-2$p$ orbitals that mainly receive
electrons on the right.

In accordance with these findings, we observe that electrostatic
screening reduces the restoring force on the displaced Ti, as a result
of the reduced lateral dipole-dipole interactions within the dipole
plane. As Fig.~\ref{fig:forces-BTO} shows, the decrease of the on-site
repulsive force is of the order of our ideal estimate of it (i.e.,
about 0.35~eV/\AA). Therefore, in this specific regard, screening {\em
  favors} the occurrence of the polar distortion.

Finally, let us note that we observe a more efficient screening, with
accumulation and depletion regions that tend to get narrower, upon
increasing the density of dopants (see Figs.~\ref{fig:forces-BTO} and
\ref{fig:charge-BTO}), as expected for a greater abundance of mobile
carriers.

\subsubsection{Doped BaTiO$_3$: Short-range effects, meta-screening}

Understandably, most discussions of free-carrier effects in the
ferroelectrics literature focus on the suppression of the long-range
electrostatic interactions. However, our results reveal another
important -- even dominant -- effect in the doped materials, one that
is largely independent of the doping type. It is a short-range,
screening-related effect that we term {\em meta-screening}, which
enhances the tendency of the material to display polar distortions.

Compared to the undoped ones, the doped systems exhibit
(Figs.~\ref{fig:forces-BTO} and \ref{fig:charge-BTO}) significantly
modified forces on atoms close to the dipole planes. These changes
happen concurrently with the accumulation of screening electrons and
holes (e.g., in the regions marked in Figs.~\ref{fig:forces-BTO} and
\ref{fig:charge-BTO}), and follow their variation in width as a
function of doping.  For atoms in those regions, the forces in the
undoped case had an obvious electrostatic character. But,
surprisingly, such forces become significantly stronger upon doping,
e.g., increasing by a factor of 2, from 0.15~eV/\AA\ to about
0.35~eV/\AA\, for $\rho$= $\pm$0.01~$|e|$/f.u. on the Ti's marked with
horizontal arrows in Fig.~\ref{fig:forces-BTO}(b). Since the
dipole-dipole interactions essentially vanish in the doped case, these
stronger forces have a different origin, and fall within the general
category of short-range interactions. This effect is associated to the
electrostatic screening, since it occurs in response to the spatial
modulation of the accumulated screening charge (almost irrespective of
its sign) around the dipole plane; yet, it clearly transcends the
screening of long-range dipolar couplings. We thus term it {\em
  meta-screening}, i.e., occurring along with, but beyond, normal
screening.

While a complete discussion of this meta-screening will require
further work, its central features lend themselves to simple
interpretations. For example, upon doping, the forces acting on the
apical O(3) closest to the displaced Ti [marked with arrows in
  Fig.~\ref{fig:forces-BTO}(d)] are positive and significantly smaller
than in the undoped case. Hence, it appears that we see in action the
repulsive interactions invoked above to rationalize these forces in
absence of doping. However, in the doped cases, the accumulation of
electrons in the Ti at z$\simeq$6~\AA\ may itself repel the O(3) anion
at z$\simeq$4~\AA\ and result in a smaller positive force than in the
undoped case; similarly, the accumulation of holes in the Ti at
z$\simeq -2$~\AA\ may attract the negatively charged O(3) at $z=0$ and
result in relatively small positive force acting on that oxygen. Such
considerations apply as well to the forces obtained for the Ti atoms
in the immediate vicinity of the dipole plane [marked with horizontal
  green arrows in Fig.~\ref{fig:forces-BTO}(b)]. The one on the right
is strongly populated with screening electrons; the obtained positive
force would tend to separate it from the displaced Ti, thus expanding
the lattice as required to accommodate such an electron excess. The
one on the left is in an electron-depleted region, and the obtained
positive force would tend to shrink the lattice on that
side. Interestingly, this interpretation is consistent with the
doping-driven pressure-like effects reported below.
 
Now, it is important to note that the largest effects observed, 
especially those pertaining to the Ti atoms closest to the dipole
plane, tend to favor the onset of a PD parallel to the imposed
dipoles.  Indeed, in the accumulation and depletion regions, the
computed forces are positive on the cations and negative on the
oxygens, and will yield a PD that is qualitatively similar to the FE
mode of undoped BTO. It is tempting to interpret the forces obtained
under doping as a consequence of imperfect screening, and a signature
of how the material tries to reduce the inhomogeneity in the
displacement field via a PD. However, as emphasized above, such an
electrostatic effect should be strongest in the undoped compound,
while we find the largest PD-favoring short-range forces in the doped
cases.

Hence, we conclude that the dominant mechanism causing the strongest
changes in the short-range forces under doping is a local lattice
response accommodating the screening electrons and holes.
Incidentally, the similarity between the meta-screening-induced
relaxation and BTO's soft FE mode -- both of which are essentially
characterized by the relative displacement of Ti-O(3) pairs -- is not
surprising: upon a local perturbation (i.e., our imposed dipole
planes), the lattice response will typically be dominated by the
lowest-energy distortions that become activated by the perturbation;
in our case, such distortions are the soft polar modes, which continue
to be rather low in energy in BTO even upon doping [this is obvious
from Fig.~\ref{fig:stiffness}(a), discussed below].

In summary, we have evidence for a previously unnoticed, short-range
meta-screening effect, which is a by-product of the electronic
screening and favors polar distortions for both $n$ and
$p$ doping. As shown below, meta-screening occurs in all the
considered perovskite oxides, hence it is likely to be a general
phenomenon.

\subsubsection{Soft modes under doping}

To address the (in)stability of cubic BTO against polar distortions
and its dependence on doping, we compute the force-constant matrix at
the $\Gamma$ point (Brillouin zone center) via standard
finite-displacement methods in our 1$\times$1$\times$20 supercell. We
focus on the $z$-polarized instability, and displace the atoms by
0.01~\AA\ from their ideal cubic positions. The $\Gamma$-point
force-constant matrix is trivially derived from the computed forces by
a supercell average. While the same $\Gamma$-point matrix can be
easily obtained in the five-atom BTO unit cell, using the long supercell
we can monitor the various interactions in real space, and modify them
by hand to test their individual effects. Note also that this
force-constant matrix yields the zone-center dynamical matrix just by
introducing suitable mass factors. Any soft-mode instability of the
cubic structure results in both matrices having (at least) one
negative eigenvalue, corresponding to a negative force constant
(energy curvature) in the former case, and to an imaginary frequency
in the latter.

\begin{figure}
\includegraphics[width=0.95\columnwidth]{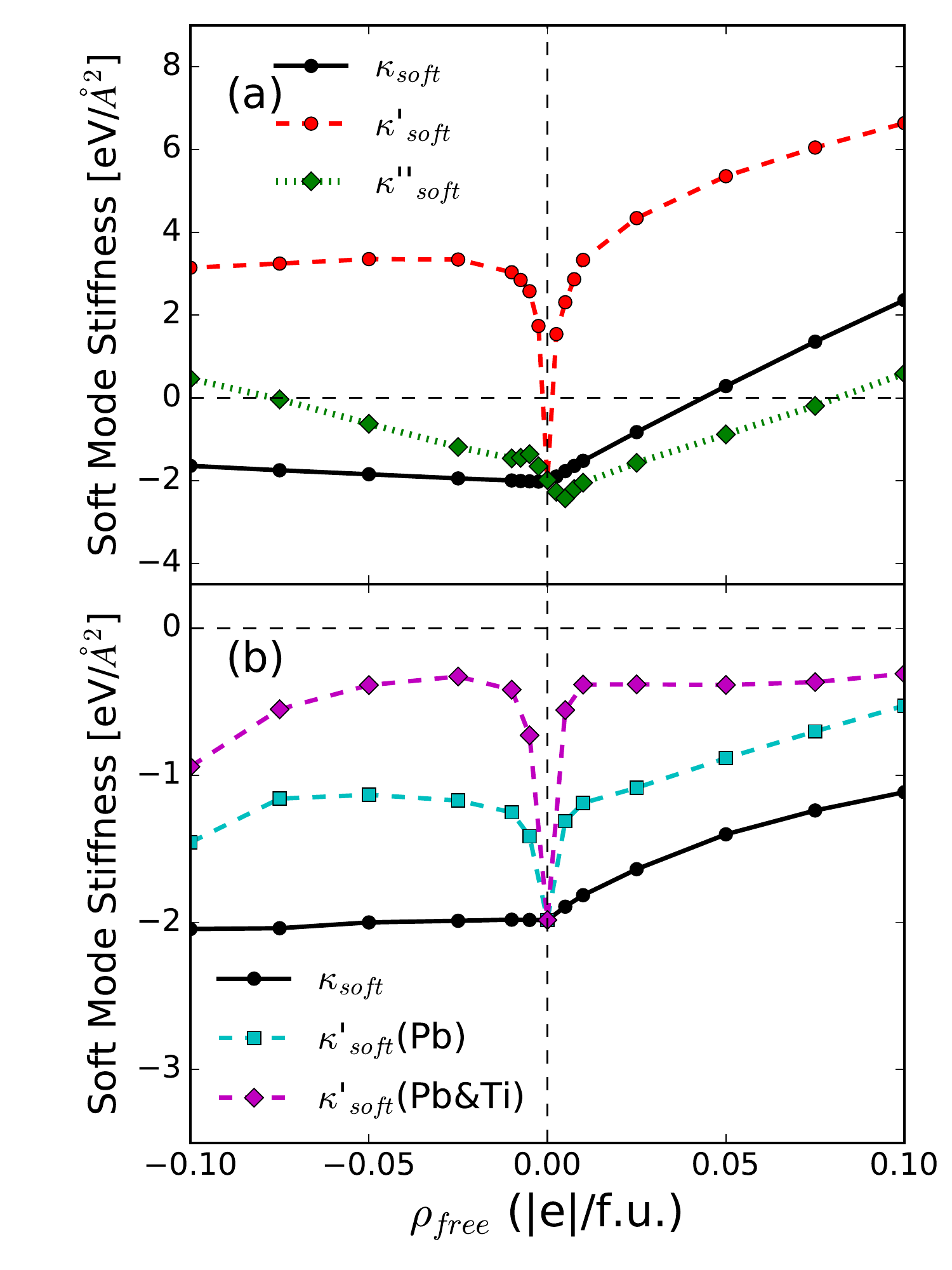}
  \caption{Ferroelectric soft-mode stiffness obtained from the
    diagonalization of the $\Gamma$-point force-constant matrix, as a
    function of doping. (a), (b) Show results for
    BaTiO$_{3}$ and PbTiO$_{3}$, respectively. The actual results are
    shown with solid lines ($\kappa_{\rm soft}$), while the results
    obtained after modifying selected interactions ($\kappa'_{\rm
      soft}$, $\kappa''_{\rm soft}$) are displayed using dashed and
    dotted lines. See text for details.}
  \label{fig:stiffness}
\end{figure}

Figure~\ref{fig:stiffness}(a) shows our basic result, i.e., the
evolution of the force constant (or stiffness) of the soft polar mode,
$\kappa_{\rm soft}$, as a function of doping. As expected, we find
that electron doping eliminates the polar instability at $\rho_{\rm
  free}\approx 0.045$~$|e|$/f.u., which roughly agrees with the
results in Fig.~\ref{fig:fe-vs-doping}. (Slight quantitative
  differences are due to volume effects, because in
  Fig.~\ref{fig:stiffness} we work with the optimized undoped cubic
  cell, while in Fig.~\ref{fig:fe-vs-doping} we optimize the cell of
  the polar structure.) In contrast, the polar instability survives
when the doping is with holes. Let us stress that our supercell
calculations only involve displacements of atoms in the unit cell at
the origin, so the settings are identical (except for the use of
smaller displacements, to make sure we are in the harmonic regime) to
those used in the dipole-plane simulations described above. Hence, all
the electronic effects discussed earlier in this paper are obviously
active in the simulations, and contribute to the obtained evolution of
$\kappa_{\rm soft}$.

We have seen above that the long-range dipole-dipole interactions --
well, established to be the driving force for ferroelectricity in
undoped BTO, are all but gone as soon as some dopants are introduced
in the material. It is thus surprising that doped BTO retains a polar
soft mode in some doping ranges. Incomplete electrostatic screening
might be a tempting explanation for the case of small $n$ doping, but
it most certainly does not apply to the results for large
$p$ doping. Instead, it seems more reasonable to turn our attention to
the meta-screening effects revealed above as a possible origin for the
observed behavior. Let us focus on the most obvious one, i.e., the
strong coupling between first-nearest-neighboring Ti atoms that
renders the very large forces marked with green horizontal arrows in
Fig.~\ref{fig:forces-BTO}(b). To test whether such an interaction may
explain the polar instability in doped BTO, we run the following
computational experiment.

The $\Gamma$-point force-constant matrix $\phi_{ij}$ and the soft
polar mode $\hat{u}_{{\rm soft},i}$ obtained from its diagonalization
satisfy
\begin{equation}
  \kappa_{\rm soft} = \sum_{ij} \hat{u}_{{\rm soft},i} \phi_{ij}
  \hat{u}_{{\rm soft},j} \, 
\end{equation}
where $i$ and $j$ run over the atoms in the unit cell and spatial
directions, and $\kappa_{\rm soft}$ is the soft-mode force constant,
depicted in Fig.~\ref{fig:stiffness}(a). Naturally, all these
quantities depend implicitly on $\rho_{\rm free}$. We now test how the
stiffness constant of the soft mode changes if we modify some key
interactions. To do this, we construct a new force-constant matrix
$\phi'_{ij}$ that is identical to $\phi_{ij}$ except that we impose
the coupling between first-nearest-neighboring Ti atoms be always that
of the undoped case, independently of the doping level. We thus remove
the most prominent meta-screening effect revealed above. The modified
stiffness
\begin{equation}
  \kappa'_{\rm soft} = \sum_{ij} \hat{u}_{{\rm soft},i} \phi'_{ij}
  \hat{u}_{{\rm soft},j} 
  \label{eq:kappa-prime}
\end{equation}
is shown as function of doping in Fig.~\ref{fig:stiffness}(a) (dashed
red lines). It is obvious that once the meta-screening effect is
removed, BTO instantly loses its polar instability upon doping,
irrespective of the sign of the extra charges. Hence, the
meta-screening effect is the driving force for the polar instability
of doped BTO.

Note that in the past, e.g., in the important work of
Wang {\em et al}. \cite{wang12}, short-range forces have
generally been assumed to be independent of doping. Based on this
(incorrect) assumption, it is most natural to attribute the
persistence of the PD in metallized BTO to the action of screened,
but strong enough, Coulomb interactions. Our present results clearly
show that this is not the case.

There is a clear $p$-$n$ asymmetry in Fig.~\ref{fig:stiffness}(a),
evidenced e.g. by the slope discontinuity of $\kappa_{\rm soft}$
around $\rho_{\rm free} = 0$. This is a direct consequence of the
existence of a band gap in the material, and of the different
character of the states occupied by the doping electrons (Ti's 3$d$)
and holes (O's 2$p$). Further, while the meta-screening effect is
sufficient to preserve the polar instability in $p$ doped BTO in this
range, it is overcome by some other interaction in the $n$ doped
compound, where the PD eventually disappears ($\kappa_{\rm soft}>0$
for $\rho_{\rm free} > 0.045$~$|e|$/f.u.). The largest and most
relevant differences between $n$ and $p$ doping do not pertain to
electrostatic screening, which is very efficient in both cases and
causes similar meta-screening effects. Instead, the greatest
differences pertain to the shortest-range interactions; most
importantly, the results in Fig.~\ref{fig:forces-BTO} show that the
restoring forces are systematically weaker for $p$-doping.

\begin{figure*}[ht]
\includegraphics[width=\textwidth]{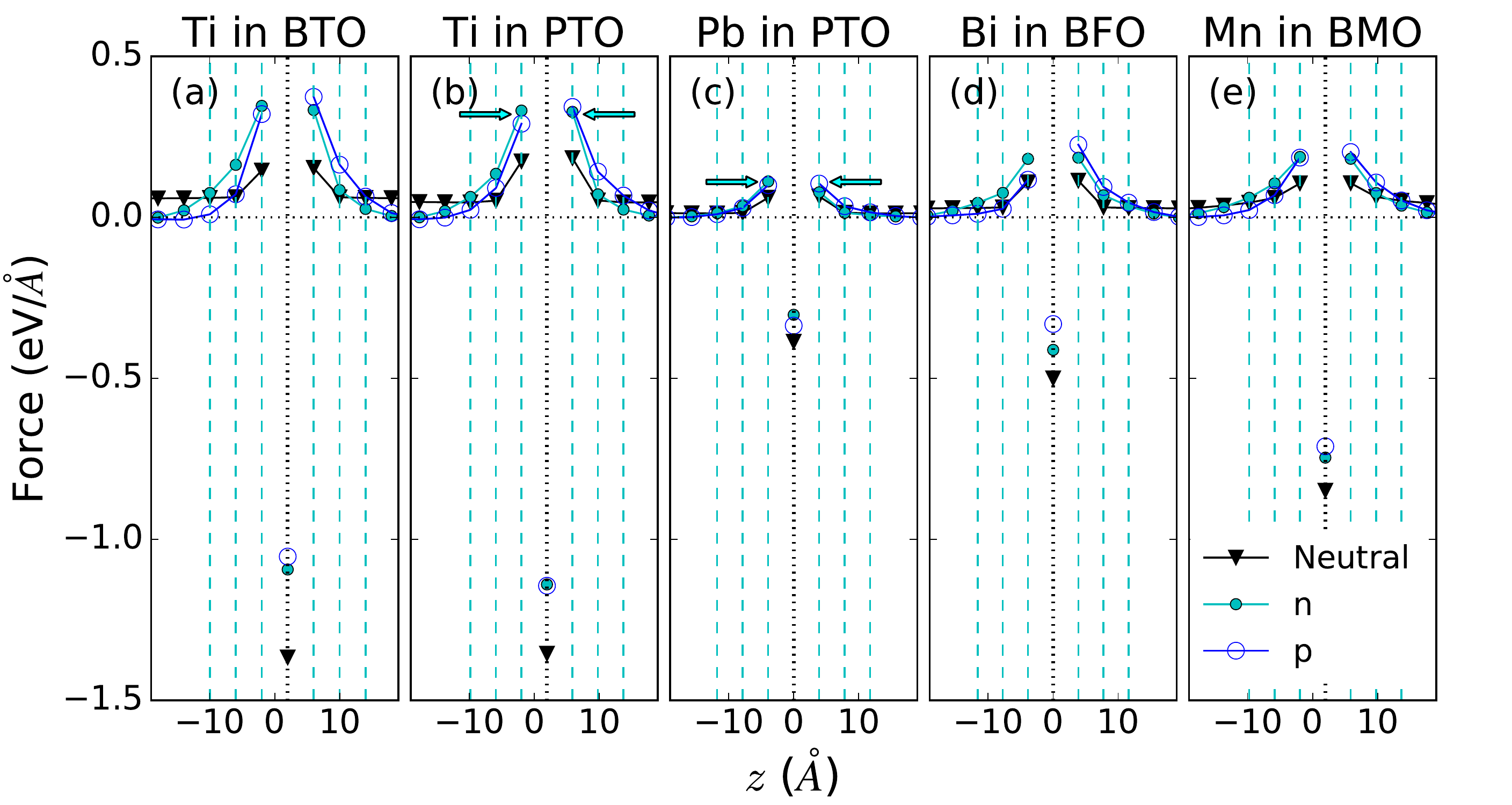}
  \caption{Same as Fig.~\ref{fig:forces-BTO}, but for other compounds
    and atoms. Panel~(a): BaTiO$_{3}$, dipole plane creating by
    displacing the Ti atoms at $z \approx 2$~\AA, forces on Ti atoms
    shown. Panel~(b): PbTiO$_{3}$, displaced Ti atoms at $z \approx
    2$~\AA, forces on Ti atoms shown. Panel~(c): PbTiO$_{3}$,
    displaced Pb atoms at $z = 0$, forces on Pb atoms
    shown. Panel~(d): BiFeO$_{3}$, displaced Bi atoms at $z = 0$,
    forces on Bi atoms shown. Panel~(e): BaMnO$_{3}$, displaced Mn
    atoms at $z \approx 2$~\AA, forces on Mn atoms shown. $n$-doping and
    $p$-doping cases correspond to $\rho_{\rm free}$ values of
    $0.01$~$|e|$/f.u. and $-0.01$~$|e|$/f.u., respectively.}
  \label{fig:forces-others}
\end{figure*}

This result can be understood by recalling the usual picture of the
Ti--O electronic hybridizations in BTO, which emphasizes the key role
of second-order Jahn-Teller effects to permit the FE distortion of
this material. In essence, the energy of the compound can be reduced
by the hybridization of (empty) Ti-3$d$ and (occupied) O-2$p$ states,
which is prompted by the onset of the PD and associated reduction of
the Ti--O(3) distance. Additional electrons would tend to occupy the
empty orbitals above the band gap, and thus increase the energy
significantly; in contrast, additional holes would occupy filled
valence states, and result in a relatively moderate energy
increase. Hence, it naturally follows that short-range restoring
(repulsive) forces will be stronger for the $n$-doping case, which is
consistent with the observed suppression of the PD only upon electron
doping.

To test the effect of these different forces, we run another
computational experiment along the lines of the one just described. We
construct modified force-constant matrices $\phi''_{ij}$ in the
following way: For a certain $n$-doping ($p$-doping) given by
$\rho_{\rm free}$, we substitute the self-interaction of the Ti atom
[responsible for the largest restoring force, marked with a gray arrow
  in Fig.~\ref{fig:forces-BTO}(b)] by the value obtained for the
corresponding $p$ doping ($n$ doping). We thus obtain a second
modified stiffness $\kappa''_{\rm soft}$; the results are in
Fig.~\ref{fig:stiffness}(a), green dotted lines. We observe a notable
degradation of the polar instability under $p$-doping, and a sizable
strengthening upon $n$-doping. (The irregular behavior of
$\kappa''_{\rm soft}$ near $\rho_{\rm free} = 0$ reflects the
qualitatively different effects of $n$- and $p$-doping on the
short-range interactions, due to the band gap. Similarly,
  the occurrence of a minimum of $\kappa''_{\rm soft}$ for $\rho_{\rm
    free} \neq 0$ is a by-product of the artificial way in which we
  construct $\phi''_{ij}$, and not worth discussing.) These results
thus indicate that the main difference between electron and hole
doping lies in their effect on the short-range repulsive couplings.

\subsubsection{Other materials}

Having discussed in detail BTO's case, our findings for BMO, PTO and
BFO are easy to present. Figure~\ref{fig:forces-others} summarizes the
results from our supercell simulations with imposed dipole planes,
which we create by displacing Ti and Pb atoms in the case of PTO
[Figs. 8(b) and (c), respectively], Bi atoms in the case of BFO
[Fig. 8(d)], and Mn atoms in the case of BMO [Fig. 8(e)]. We also
include in Fig. 8(a) the results for BTO, for an easier
comparison. Remarkably, the computed forces exhibit the same essential
features discussed above for BTO.

Most importantly, we emphasize that meta-screening, i.e. the
enhancement of short-range interactions upon doping, occurs in all the
considered materials, and is thus very likely to be a general
phenomenon. Moreover, in all cases, meta-screening favors again polar
distortions. (Figure~\ref{fig:forces-others} shows positive forces on
the key cations; the forces on the oxygens, not shown here, are
negative.)

To drive this point home, we show in Fig.~\ref{fig:stiffness}(b) three
versions of the stiffness constant of the soft mode of PTO as a
function of doping. Similarly to BTO, we present the stiffness
$\kappa_{\rm soft}$ obtained from the $\Gamma$-point force-constant
matrix, along with two other quantities: one is $\kappa'_{\rm
  soft}$(Pb), obtained from Eq.~(\ref{eq:kappa-prime}) for the same
matrix, except for the strongest meta-screening forces acting on Pb
ions [marked with arrows in Fig.~\ref{fig:forces-others}(c)] being
replaced by the corresponding values in the undoped case. If we also
similarly modify the forces acting on the Ti ions [marked with arrows
  in Fig.~\ref{fig:forces-others}(b)] we obtain by the same procedure
a third stiffness variant, $\kappa'_{\rm soft}$(Pb\&Ti). Essentially,
when the system is purged of the meta-screening couplings, the soft
modes are much less soft, i.e., their force constants are much less
negative, in accordance with our previous conclusion that
meta-screening is the main driver of the permanence of PDs in doped
FEs.

We should note that, from the evidence at hand, we cannot tell whether
the meta-screening mechanism is a necessary condition for the PD to
occur in a compound like PTO. To elucidate that question, we would
need an accurate quantification of the meta-screening contribution to
the forces, so that such effects can be clearly disentangled from
other (steric/chemical) factors. This poses an interesting and
non-trivial challenge to electronic-structure theory, and remains for
future work.

The results in Fig.~\ref{fig:forces-others} offer other interesting
insights. For example, it is apparent that the restoring forces are
relatively small for the Pb$^{2+}$ (in PTO) and Bi$^{3+}$ (in BFO)
cations, and relatively large for Ti$^{4+}$ (in both BTO and PTO) and
Mn$^{4+}$ (in BMO). We think this difference can be partly attributed
to the stereochemical activity of Pb$^{2+}$ and Bi$^{3+}$'s lone
pairs, which tends to compensate the electronic repulsion between
ionic cores.

It is also interesting to note that the restoring force acting on the
displaced Mn$^{4+}$ (3$d^3$) cation in BMO is significantly smaller
than that on displaced Ti$^{4+}$ (3$d^0$) cation in both BTO and
PTO. This may seem at odds with the usual view that empty 3$d$
orbitals are indispensable for {\sl B}-site driven ferroelectricity to
occur. Yet, one should note that, as regards the possibility that a
Mn$^{4+}$ cation in an O$_{6}$ environment drives ferroelectricity,
the most relevant 3$d$ orbitals are those with $e_{g}$ symmetry, which
are directed towards the oxygen anions and are empty in this
case. Hence, ferroelectricity in BMO should not be penalized by strong
repulsive forces associated to the Mn$^{4+}$-3$d^{3}$ configuration
\cite{filippetti02,bhattacharjee09}. Having said this, to explain why
the restoring forces acting on BMO's Mn$^{4+}$ cation are
significantly smaller than those obtained for BTO's Ti$^{4+}$, we
probably should resort to simple steric arguments. Indeed, the ionic
radii of Ti$^{4+}$ and Mn$^{4+}$ in an octahedral O$_{6}$ environment
are 0.605~\AA\ and 0.53~\AA, respectively \cite{shannon76}; then,
noting that BTO and BMO share the same {\sl A}-site cation, size
considerations suggest that it will be easier for the smaller
Mn$^{4+}$ to move off-center, which is clearly consistent with the
relatively weak restoring force obtained in our
calculations.

Finally, let us remark the striking similarity between our results for
the Ti forces in BTO [Fig.~\ref{fig:forces-others}(a)] and the
corresponding ones in PTO [Fig.~\ref{fig:forces-others}(b)]; this
suggests that interactions between same atom pairs are relatively
unaffected by the different chemical environment in different
perovskite oxides, an observation that is in line with previous
first-principles studies \cite{ghosez99}. Additionally, note that the
results for the Pb forces in PTO [Fig.~\ref{fig:forces-others}(c)] and
the Bi forces in BFO [Fig.~\ref{fig:forces-others}(d)] are quite
similar as well. While we do not want to overinterpret these
observations, they are clearly suggestive of the hybrid nature of
ferroelectricity in PTO, as the polar soft mode of this material is
obviously participated by both the {\sl A} and {\sl B} cationic
sublattices; in contrast, BFO and BTO are textbook examples of
compounds in which ferroelectricity is driven by only one cation
sublattice, respectively {\sl A} and {\sl B}.

\subsection{Additional remarks}

\subsubsection{Volume changes and transitions under doping}

\begin{figure}
\includegraphics[width=\columnwidth]{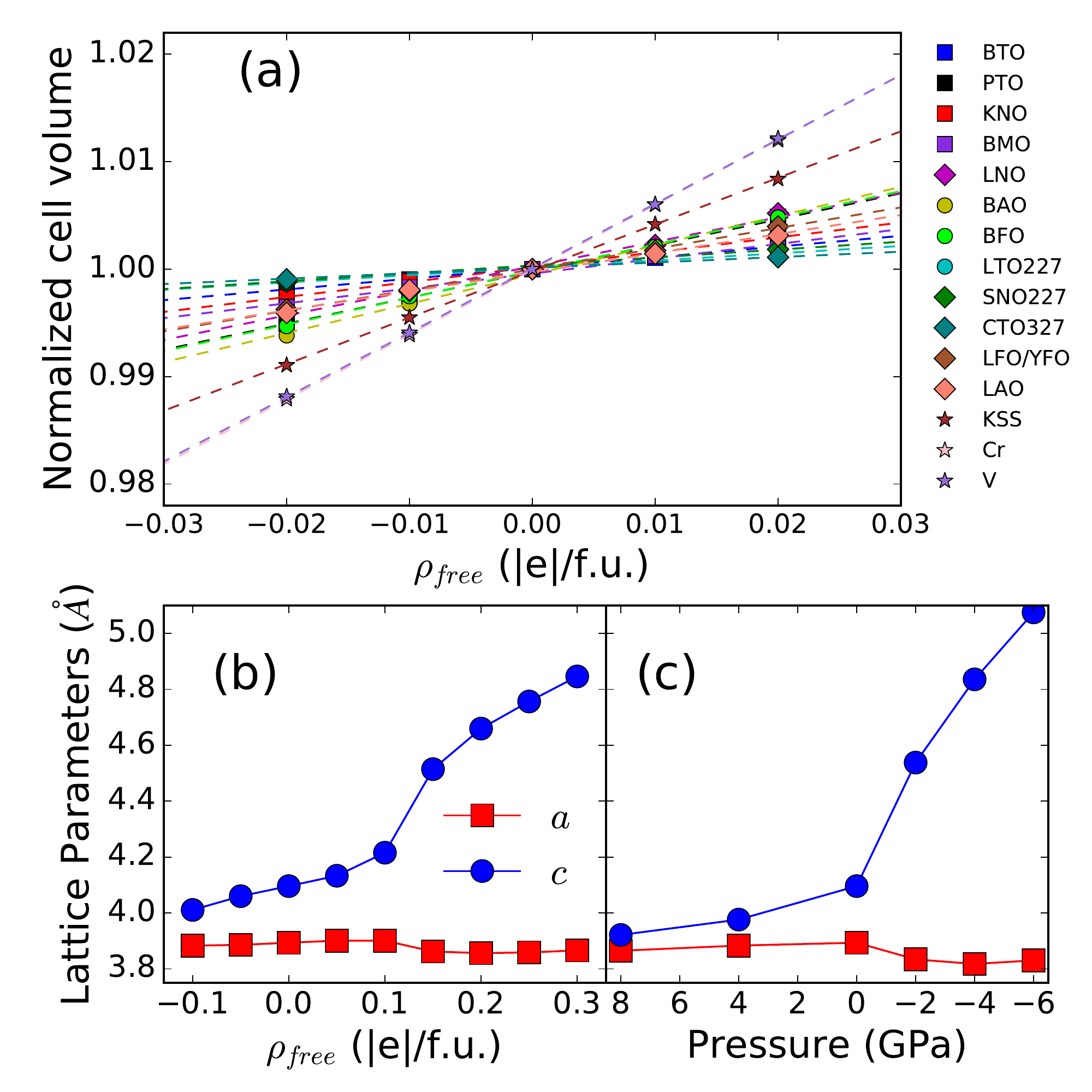}
  \caption{(a) Shows the variation of the unit cell volume,
    normalized to the $\rho_{\rm free} = 0$ result, as a function of
    doping for the 11 FE materials considered in this work, as well as
    LaAlO$_{3}$ and three other compounds (KSS, Cr and V) studied for
    comparison. The slope is positive in all cases, varying from
    0.05~f.u.$/|e|$ for CTO327 to 0.60~f.u.$/|e|$ for Cr. (b)
    Shows the evolution of the lattice constants ($a = b$ and $c$) of
    the five-atom tetragonal cell of PbTiO$_{3}$ as a function of
    doping. A transition to a super-tetragonal phase with $c \gg a$
    occurs at $\rho_{\rm free} \approx 0.125 |e|$/f.u. (c) Shows
    the analogous results, but obtained this time for undoped
    PbTiO$_{3}$ ($\rho_{\rm free} = 0$) as a function of an external
    hydrostatic pressure. The transition to the super-tetragonal phase
    occurs at $p \approx -1$~GPa.}
  \label{fig:volume}
\end{figure}

As shown in Fig.~\ref{fig:volume}(a), our simulations yield a
universal behavior regarding the volume of the doped materials:
additional electrons cause an expansion, while additional holes cause
a contraction. Such an effect had already been observed in the past,
in independent investigations of BaTiO$_{3}$ \cite{iwazaki12},
BiFeO$_{3}$ \cite{he16a} and PbTiO$_{3}$ \cite{he16b}. Our present
work confirms this behavior and shows that it pertains to all the
diverse ferroelectrics here considered.

One may wonder whether this volume effect has any influence on the
survival, or disappearance, of the PD upon doping. To check this, in
Fig.~\ref{fig:fe-vs-doping} we compared the results obtained for
constant volume [Fig. 1(a)] and relaxed volume [Fig. 1(b)], noting
that in the considered doping range the volume changes can be up to
$\pm 4$~\%. Our results show that FEs conserve their PD irrespective
of whether we allow the volume to relax or not (with the partial
exception of $n$-doped BTO, KNO and BMO). This suggests that the
effects discussed above, responsible for the disappearance (screening)
or survival (meta-screening) of the PD, are not much affected by even
fairly substantial volume changes.

Naturally, we do find some differences when volume relaxation is
allowed. For example, it is apparent that the contraction associated
to $p$ doping is detrimental to the PD of BTO and BMO. This result
lends itself to a simple interpretation, as it is well-known that a
compression tends to weaken ferroelectricity in conventional
perovskite oxides like BTO \cite{ishidate97,iniguez02}.

As emphasized by other authors \cite{he16a}, the doping-driven volume
changes operate in essentially the same way as a hydrostatic pressure
would, and can potentially induce structural phase transitions beyond
those (polar to non-polar) discussed above. As an example, in
Fig.~\ref{fig:volume} we show the behavior of PTO under $n$ doping and
under a negative pressure [Figs. 9(b) and (c), respectively]. In both
cases, the volume increase causes a transition into a so-called
super-tetragonal phase with giant $c/a$ aspect ratio
\cite{tinte03,wang15}. The analogy between doping and pressure is
further ratified by our studies of BiFeO$_{3}$ and LaAlO$_{3}$ [see
Supplemental material (Note 2 and Figs.~2--4) \cite{SupMat}], and suggests that non-trivial
structural effects may occur, to some extent at least, whenever
dopants stay spatially delocalized.

We can try to rationalize the volume changes in terms of the
bonding/anti-bonding character of the electronic states affected by
the doping. As described in the Supplemental Material (Note 3 and Figs.~5--12) \cite{SupMat},
some of our results are straightforwardly interpreted (e.g.,
$n$ dopants occupy anti-bonding states in our insulating oxides, which
suggests a lattice expansion consistent with our calculations), and
others can be explained by invoking plausible second-order orbital
mixing effects. Yet, we also find examples (in particular, for the
non-oxidic materials V, Cr and KSS) where such bonding arguments
clearly fail, which questions their general validity. We are thus
inclined to believe that the obtained volume effects may be the
consequence of a rather crude {\sl steric} mechanism of sorts (grossly
speaking: electrons do occupy space), which prevails over the bonding
characteristics of the (de)populated states.

We also note that our way of simulating doping is not expected to
reproduce polarons. Since previous work suggests that in some cases
volume changes are suppressed when chemical dopants \cite{iwazaki12}
or self-trapped electrons and holes \cite{he16a} are considered
explicitly, the doping-driven volume changes just reported should be
considered realistic insofar as the free charges remain
extended. Since localization is frequent in oxides, our volume changes
may be considered an upper limit when compared with experiment, but
should apply fairly closely when the injected charge is delocalized,
as at metal/ferroelectric interfaces (where some charge spillage
always occurs) and in the case of field-effect injection or
electrostatic doping.

\subsubsection{Hyperferroelectrics}

Hyperferroelectric compounds \cite{garrity14} are soft-mode
ferroelectrics whose paraelectric phase displays an unstable
longitudinal-optical (LO) polar phonon band. To obtain such an exotic
property, which suggests, e.g., that an hyperferroelectric can form
(meta)stable FE domain walls that would be formally charged, it is
mandatory to have unstable transversal-optical (TO) polar phonons and
a relatively small LO-TO splitting. The latter is typical of materials
with large high-frequency dielectric permittivity $\epsilon_{\infty}$,
i.e., materials with a very efficient electrostatic screening. Hence,
whenever we have a hyperferroelectric that displays regular (TO) FE
instabilities in spite of weak dipole-dipole interactions, that is a
good candidate to remain polar when such couplings are totally
screened ($\epsilon_{\infty}$ diverges upon doping). Conversely,
materials that remain polar upon metallization may in principle be
good candidates for hyperferroelectricity.

To investigate this connection, we looked for hyperferroelectricity in
a subset of our considered FE materials, by running straightforward
phonon and perturbative calculations that allow us to compute the
LO-TO splitting [see details in Supplemental Material (Note 4 and Table I) \cite{SupMat}]. To
our surprise, we find that only four compounds (LNO, LTO227, SNO227
and CTO327) are hyperferroelectric, while most of the materials
displaying a strong and robust PD upon doping are not. Indeed, in
materials like PTO and BFO, while the zone-center (TO) polar
instability of the cubic phase is very strong, the LO-TO splitting is
even stronger, yielding a stable LO band. Note that the very large
LO-TO splitting that is typical of FE perovskite oxides can be traced
back to the anomalously large polarity of the soft modes (which in
turn reflects unusually large dynamical charges \cite{zhong94b}) and
their relatively small $\epsilon_{\infty}$.

\section{Conclusions}

In conclusion, our first-principles study of diverse ferroelectrics
shows that their characteristic polar distortion is generally stable
upon charge doping. Remarkably, our results reveal a previously
unnoticed {\em meta-screening} effect that is essential to the
permanence of the non-centrosymmetric phase. This seemingly-universal
meta-screening mechanism is triggered by the rearrangement of mobile
electrons and holes associated to the screening of dipolar
interactions, is essentially independent of the sign of the doping
charges, and results in short-range couplings favoring a polar
distortion. Our results thus provide unprecedented insight into the
behavior of metallized ferroelectrics, potential implications ranging
from the discovery of new polar metals to the design of
metal/ferroelectric interfaces or charge-injection effects in these
compounds.

\acknowledgments

Work supported by the Luxembourg National Research Fund through Grants
P12/4853155 COFERMAT (H.J.Z. and J.I.), INTER/MOBILITY/15/9890527
GREENOX (L.B., H.J.Z. and J.I.), and AFR Grant No. 9934186
(C.E.S.). Additionally, V.F. was supported by Progetto biennale di
ateneo UniCA/FdS/RAS 2016; E.C. by the Spanish MINECO through the
Severo Ochoa Centers of Excellence Program under Grant SEV-2015-0496,
as well as through Grant FIS2015-64886-C5-4-P, and by Generalitat de
Catalunya (2017SGR1506); and L.B. by the Air Force Office of Scientific
Research under Grant No. FA9550-16-1-0065. Computational resources
have been provided by the PRACE-3IP DECI-13 grant 13DECI0270 INTERPHON
(Salomon cluster at the Czech National Supercomputing Center), the
CRS4 Computing Center (Piscina Manna, Pula, Italy), and the Arkansas
High Performance Computing Center. Also, we are grateful to M. Stengel
and P. Zubko for fruitful discussions.

\appendix\section{\label{sec:methods}Methods}

We use density functional theory (DFT) within the generalized gradient
approximation (PBEsol functional \cite{perdew08}) as implemented in
the software package {\sc VASP} \cite{kresse96,kresse99}. For all
considered compounds, the electronic wave functions are represented in
a basis of plane waves truncated at 500~eV. Reciprocal space integrals
are computed using $k$-point grids that are equivalent to (or denser
than) a $12\times12\times12$ sampling of the Brillouin zone of an
elemental five-atom perovskite cell. The interaction between ionic cores
and electrons is treated within the so-called plane augmented wave
(PAW) approach \cite{blochl94}, solving explicitly for the following
electrons: O's 2$s$ and 2$p$; Li's 2$s$; K's 3$s$, 3$p$, and 4$s$;
Ba's 5$s$, 5$p$ and 6$s$; Pb's 6$s$ and 6$p$; Ca's 3$p$ and 4$s$; Sr's
4$s$, 4$p$, and 5$s$; Bi's 6$s$ and 6$p$; La's 5$s$, 5$p$, 5$d$, and
6$s$; Y's 4$s$, 4$p$, 4$d$ and 5$s$; Al's 3$s$ and 3$p$; Ti's 3$d$ and
4$s$; Mn's 3$d$ and 4$s$; Fe's 3$d$ and 4$s$; Nb's 4$s$, 4$p$, 4$d$,
and 5$s$; Sn's 5$s$ and 5$p$; Sb's 5$s$ and 5$p$; Cr's 3$d$ and 4$s$;
and V's 3$d$ and 4$s$. For Fe's 3$d$ electrons we use the ``Hubbard
correction'' introduced by Dudarev {\sl et al}. \cite{dudarev98} with
$U_{\rm eff} = 4$~eV; for Mn's 3$d$ electrons we use the correction
introduced by Liechtenstein {\sl et al}. \cite{liechtenstein95} with
$U = 4$ eV and $J = 1$ eV. (In the case of BiFeO$_{3}$,
  we explicitly verified that our results for the persistence of the
  PD upon doping remain essentially the same for $U_{\rm eff}$ values
  between 3 and 5~eV.) Structural relaxations are run until
residual forces and stresses fall below 0.005~eV/\AA\, and 0.05~GPa,
respectively. These calculations conditions were checked to render
sufficiently converged results.

We simulate the effect of doping by varying the number of electrons in
the cell, and adding a neutralizing homogeneous charge
background. This approach, the standard one employed in most of the
previous works on this problem
\cite{iwazaki12,wang12,benedek16,he16a,he16b}, does not describe the
doping species explicitly, which greatly simplifies the
calculation. Further, we use the smallest cells describing the
equilibrium structures of the undoped material, namely a 5-atom cell
for perovskites like BaTiO$_{3}$ and PbTiO$_{3}$, a 10-atom cell for a
material like BiFeO$_{3}$, etc. Such settings impose restrictions on
the possible arrangements of added electrons or holes, such as for
example polaron states (we note in passing that standard semi-local
density functional methods are {\em a priori} not expected to yield
stable states of that type). We thus expect that our simulations will
tend to exaggerate the tendency towards metallization
  and the effectiveness of doping in producing screening, as well as 
in modifying the structure. Nevertheless, as evidenced by
  the results here reported, these idealized conditions are relevant
  to better understand the intrinsic response of FE materials to
  carrier doping. On the other hand, our results are directly 
relevant to situations that are typical of ferroelectric
nanostructures, e.g., whenever the ferroelectric material is partly
metallized near the interface with an electrode, or extra carriers are
injected by electrostatic doping, etc.

For the ferrites (BiFeO$_{3}$ and LaFeO$_{3}$/YFeO$_{3}$) and
manganite (BaMnO$_{3}$), we use the well-known lowest-energy spin
arrangement (anti-ferromagnetic with anti-parallel nearest-neighboring
spins) and the standard scalar-magnetism (collinear)
approximation. Note that, according to previous studies
  \cite{wojdel09, wojdel10}, non-collinear magnetism and spin-orbit
  interactions are expected to have a negligible impact on the FE
  instabilities of these compounds; hence, we do not consider them
  here.

We use standard analysis tools to study the doping-induced effects. In
particular, we use the FINDSYM \cite{stokes05} and AMPLIMODES 
\cite{orobengoa09,perez-mato10} codes to determine the space group of
our doped structures and to calculate the mode-resolved distortion
amplitudes, respectively. When computing the distortion amplitudes
with AMPLIMODES, the undoped high-symmetry phase ($Pm\bar{3}m$ for
simple perovskites, $I4/mmm$ for layered perovskite
Ca$_{3}$Ti$_{2}$O$_{7}$, $Cmcm$ for layered perovskites
La$_{2}$Ti$_{2}$O$_{7}$ and Sr$_{2}$Nb$_{2}$O$_{7}$, and $P4/mbm$ for
superlattice LaFeO$_{3}$/YFeO$_{3}$) is taken as the reference
structure. Note that when AMPLIMODES compares a reference CS structure
with a polar one (doped or undoped), it will in general yield a
collection of amplitudes corresponding to modes of different
symmetries; from those, we retain the result corresponding to the
polar mode (which e.g. corresponds to the $\Gamma_{4}^{-}$ irreducible
representation in the case of simple perovskites) to quantify the
CS-breaking distortion.

Finally, we also use the {\sc ASE} tools \cite{ase-paper,matplotlib}
and {\sc VESTA} \cite{momma11} for analysis and visualization of our
results, as well as the LOBSTER code
\cite{dronskowski93,deringer11,maintz13,maintz16a,maintz16b} to
characterize the bonds and electronic structure via a standard COHP
(crystal orbital hamilton population) analysis.


%

\end{document}